\documentclass[10pt,journal,compsoc,twoside]{IEEEtran}
\usepackage{amsmath,amsfonts}	% text
\usepackage{booktabs}			% toprule
\usepackage{array}				% table
\usepackage{hyperref}			% href
\usepackage{cite}
\usepackage{graphicx}
\usepackage{adjustbox}
\usepackage{multirow}
\usepackage[caption=false,font=footnotesize,labelfont=rm,textfont=rm]{subfig}
\usepackage[linesnumbered,ruled,boxed]{algorithm2e}
\usepackage{mdframed}
\definecolor{bottomcolor}{RGB}{247, 248, 236}

\usepackage{ragged2e}
\renewcommand{\justify}{
	\leftskip=0pt \rightskip=0pt plus 0cm}

\hypersetup{
	colorlinks=true,
	linkcolor=black,
	citecolor=black,
	urlcolor=blue,
}

\begin{document}

%\markboth{IEEE TRANSACTIONS ON PATTERN ANALYSIS AND MACHINE INTELLIGENCE}%
{}

\title{\LARGE A New Brain Network Construction Paradigm for \\ Brain Disorder via Diffusion-based Graph Contrastive Learning}

\author{Yongcheng Zong, Shuqiang Wang
	\thanks{Corresponding author: Shuqiang Wang, sq.wang@siata.ac.cn}
	\thanks{Yongcheng Zong and Shuqiang Wang are with the Shenzhen Institutes of Advanced Technology, Chinese Academy of Sciences, Shenzhen, 518055, China}
   
}

\IEEEtitleabstractindextext{%
\begin{abstract}
\justify
Brain network analysis plays an increasingly important role in studying brain function and the exploring of disease mechanisms. However, existing brain network construction tools have some limitations, including dependency on empirical users, weak consistency in repeated experiments and time-consuming  processes. In this work, a diffusion-based brain network pipeline, DGCL is designed for end-to-end construction of brain networks. Initially, the brain region-aware module (BRAM) precisely determines the spatial locations of brain regions by the diffusion process, avoiding subjective parameter selection. Subsequently, DGCL employs graph contrastive learning to optimize brain connections by eliminating individual differences in redundant connections unrelated to diseases, thereby enhancing the consistency of brain networks within the same group. Finally, the node-graph contrastive loss and classification loss jointly constrain the learning process of the model to obtain the reconstructed brain network, which is then used to analyze important brain connections. Validation on two datasets, ADNI and ABIDE, demonstrates that DGCL surpasses traditional methods and other deep learning models in predicting disease development stages. Significantly, the proposed model improves the efficiency and generalization of brain network construction. In summary, the proposed DGCL can be served as a universal brain network construction scheme, which can effectively identify important brain connections through generative paradigms and has the potential to provide disease interpretability support for neuroscience research.
\end{abstract}

}

\maketitle

\IEEEdisplaynontitleabstractindextext

% \fi
\IEEEpeerreviewmaketitle

\IEEEraisesectionheading{\section{Introduction}\label{sec:introduction}}

\IEEEPARstart{B}{rain} diseases represent a broad spectrum of conditions that significantly impact both individuals and society as a whole. These disorders such as Alzheimer's disease (AD) and autism spectrum disorder (ASD) encompass a wide range of brain-related challenges, affecting individuals across different age groups and backgrounds \cite{1}. The complexity of these disorders, coupled with their varied etiologies, makes them a formidable challenge for researchers and clinicians alike. Furthermore, the absence of precise biomarkers and the often subtle nature of early-stage symptoms make early detection a daunting task \cite{2}. Neuroimaging technologies such as magnetic resonance imaging can provide non-invasive and accurate diagnostic methods, and become an important auxiliary means to diagnose brain diseases. Magnetic resonance imaging (MRI) is a non-invasive, repeatable, and high spatial resolution technique widely used in the diagnosis and research of various brain diseases\cite{3}. Among them, diffusion tensor imaging (DTI) is a special form of magnetic resonance imaging that can be used to describe brain structures. According to the DTI reconstruction algorithm, the direction of white matter fiber bundles can be obtained, and finally the structural join of brain networks can be constructed. This technology is widely used for the effective identification of various neurodegenerative diseases and the classification of AD development stages, especially for the recognition of abnormal brain connections \cite{4}.

To address these challenges, recent years have seen the rapid development of deep learning techniques in the domain of neuroimaging analysis, which has, in turn, been marked by considerable success \cite{5,6,7}. Among them, graph neural networks (GNN) hold promise for modeling intricate, complex networks, such as brain networks. Researchers have increasingly employed GNNs to study brain networks, improving the interpretability of machine learning in brain network analysis \cite{GNN2020}. However, these methods typically require preprocessed graph structures as input, which are obtained by medical imaging software such as SPM, resulting in uncertainty of generated brain networks. In these studies, considering the differences in the target dataset and subjective selection of various parameters, there will be significant differences in the preprocessing steps. In addition, traditional software heavily relies on setting a large number of parameters, and both optimizing and selecting the correct parameters requires a lot of empirical guidance. Different parameter settings may affect the preprocessing results\cite{11}. When constructing brain networks, traditional research methods often rely on statistical analysis at the group level, mainly using graph attributes and related graph theory indicators for classification. At the same time, traditional software-based methods (such as PANDAS) also face some difficulties in non-standard individual space image registration, alignment, and other issues. Overall, due to the complex brain structure and individual differences, research on individual-level brain networks and connectivity changes remains a huge challenge.

% htbp
\begin{figure*}[!t]
	\centering
	\includegraphics[width=\textwidth]{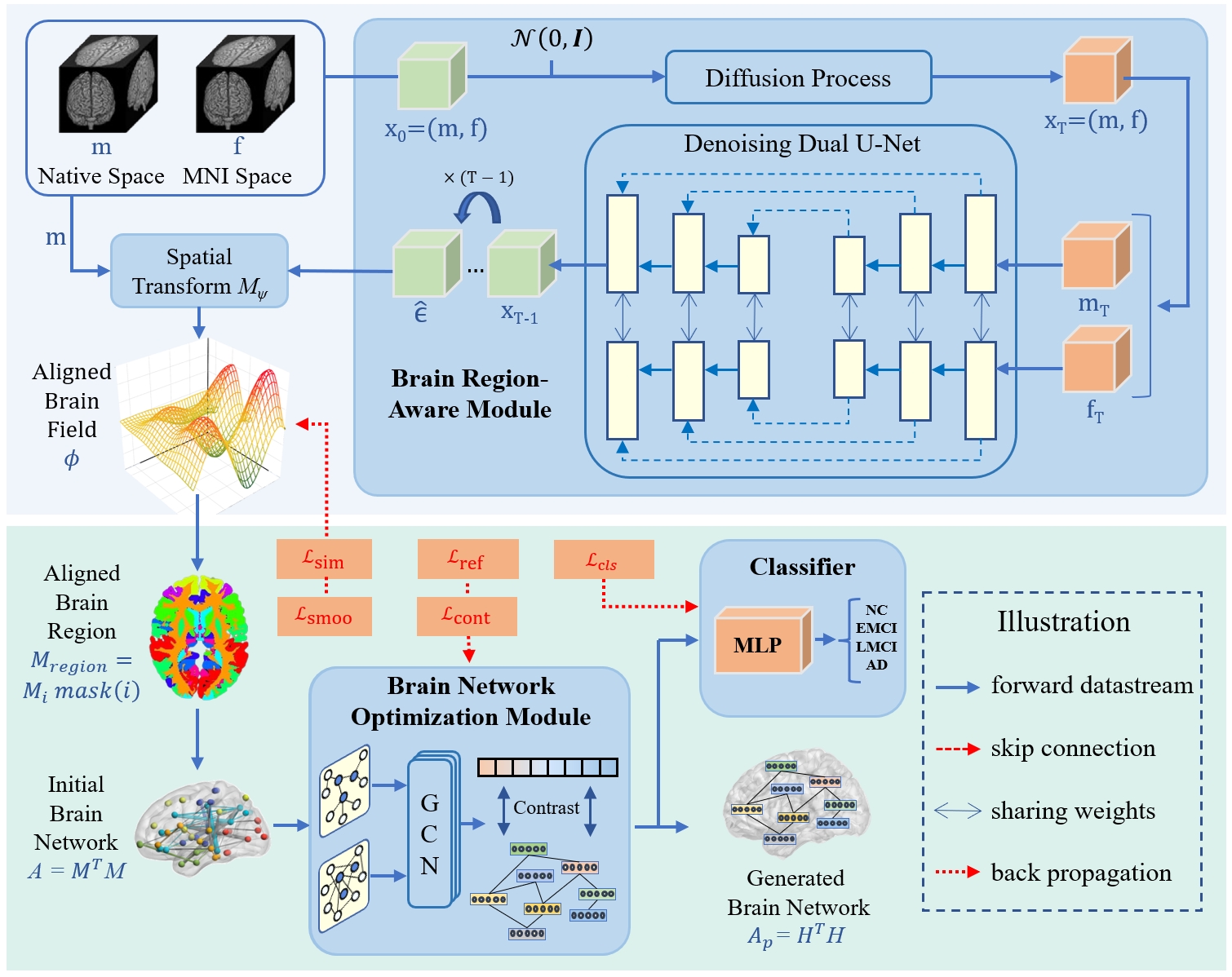}
	\caption{The network architecture of the proposed DGCL has three sub-modules: (a) Brain Region-Aware Module, (b) Brain Network Optimization Module, and (c) MLP Classifier. For the convenience of description, we only show the process of the second part and the implementation details are illustrated in Fig. 2.}
	\label{fig_1}
\end{figure*}

Compared with traditional software-based methods, the proposed method is completely data-driven and end-to-end processing flow, avoiding the tedious image data preprocessing steps on traditional software, and improving operational efficiency and accuracy. In this paper, we propose a novel deep learning-based method named DGCL for constructing brain networks, and the main contributions are summarized as follows.
\begin{enumerate}
	\item{
		A new paradigm for constructing brain networks has been proposed, which is completely an end-to-end pipeline for constructing brain networks from brain images. Compared with existing methods that rely on software templates, it can reduce technology dependence, improve the running efficiency, and generate consistent brain networks.}
	\item{
		A Brain Region-Aware Module (BRAM) based on the diffusion mechanism is proposed, which can align and register unprocessed images in non-standard native spaces. After learning more precise information regarding the divisions of brain regions, BRAM demonstrates the capacity to automatically align a wide range of brains, spanning from children to adults.}
	\item{
		The Brain Network Optimization Module (BNOM) utilizes the idea of graph contrastive learning to effectively remove redundant features such as individual differential connections that are unrelated to diseases. By retaining common connections between subjects and enhancing important brain connections, the consistency and robustness of brain network construction have been improved.
	}
\end{enumerate}

\section{Related work}
There has been an increasing amount of research on brain network construction and analysis in recent years, particularly for Alzheimer's disease diagnosis and treatment. Considering the lack of research on using diffusion models for the brain network generation, we will review the current research status from three aspects.

\subsection{Brain network analysis methods}
For the research of AD diagnosis, the most famous public neuroimaging data set comes from the Alzheimer's Disease Neuroimaging Initiative (ADNI), which is committed to improving the clinical trials of AD prevention and treatment \cite{ADNI2005}. And the largest dataset for autism research comes from Autism Brain Imaging Data Exchange (ABIDE)\cite{abide}. Neuroscientists use neural image data to model the brain as a graph network structure, where graph nodes represent anatomical brain regions connected by edges, while graph edges can be defined according to brain morphology, function or structural connectivity related to paired nodes \cite{9}. Different modalities of brain imaging data are commonly used to construct three types of brain networks: morphological networks, structural networks, and functional networks, respectively, from structural T1 weighted imaging (T1WI), diffusion-weighted imaging (DTI), and resting state functional MRI (rs fMRI). He et al. constructed the brain morphological brain network for the first time in 2007 using the cortical thickness of various brain regions as morphological features \cite{10}, and discovered the small world attribute of the brain network. Afterwards, a large number of complex network analysis methods based on graph theory began to study the connection patterns and graph attribute analysis between brain regions, which were used to explore the diagnosis of brain functional disorders such as mental disorders from a structural to functional perspective. To simplify brain network analysis, the threshold method is usually used to sparsize the network. Research has shown that sparsization operations have a significant impact on subsequent analysis \cite{11}. So far, there is still no optimal standard for determining the threshold. In experiments, cross-validation is generally used to traverse all possible value ranges to determine the optimal threshold.

\subsection{Graph neural network methods}
In recent years, with the vigorous development of deep learning theory, various advanced neural network models have been used for the analysis and processing of medical images, especially the success of graph neural networks, which provides a new research direction for traditional brain network models based on statistical analysis. Graph neural networks are one of the current research hotspots in deep learning technology, and have become the most widely used method for extracting information from graphs to perform the diagnosis of brain diseases. Ma et al. innovatively proposed an attention-guided deep map neural network in 2020 for analyzing AD data from sMRI \cite{12}, which can reveal the most relevant brain regions and basic time points of AD progression. Gan et al. proposed a multi-graph fusion method in 2021 to explore common and complementary information between functional brain networks \cite{13}. Zhu et al. proposed a dynamic graph convolutional network in 2022 and used feature learning techniques to make it easier to interpret \cite{14}. Cui et al. first proposed a benchmark for brain network analysis using message-passing graph convolutional networks in 2022 \cite{15}, providing direction for establishing a unified, scalable, and replicable GNN brain network analysis framework. However, these methods rely on software preprocessing to construct initial brain networks, requiring manual parameter settings on different modal datasets, making it difficult to replicate and fairly compare the final experimental results under the same conditions.

\subsection{Deep learning methods}
Another end-to-end learning approach attempts to automatically learn features from each voxel in brain images. For example, in the field of natural images,  convolutional neural networks(CNN) based methods learn all levels of features from the original pixels, avoiding the process of manual ROI (regions of interest) annotation. Such methods can model or focus on disease-related brain areas such as gray matter or hippocampus at the whole brain level \cite{H-FCN}. Sarraf et al. used functional magnetic resonance imaging data for deep learning applications for the first time in 2016 \cite{16}. In the preprocessing step, 4D rs-fMRI and 3D MRI data are decomposed into 2D format images, and then these images are received in their input layer based on a CNN architecture. Feng et al. realized a deep learning network based on 3D-CNN and full stack bidirectional short-term memory (FSBi LSTM) in 2019 \cite{17}. Among them, the 3D-CNN network is used to extract features, while FSBi-LSTM extracts advanced semantic and spatial information. Kang et al. borrowed from the idea of ensemble learning in 2021 and effectively improved classification accuracy and stability by integrating multiple MRI 2D slices \cite{18}. Loddo et al. conducted a comparative study on the work of deep learning in the field of AD diagnosis in 2022, and then used transfer learning strategies on different heterogeneous data sets to detect AD diseases and distinguish dementia of different degrees. The experimental results show that among the CNN network types considered, ResNet-50 and ResNet-101 models are the most appropriate solutions for transfer learning \cite{resnet, 19}. Although deep learning methods are convenient and efficient, with continuous breakthroughs in the accuracy of results, most of the existing work focuses on brain segmentation or classification tasks, making it difficult to accurately locate abnormal changes in brain regions and lacking interpretability in neuroscience.

In summary, the above studies have made significant contributions to the field of brain network construction and analysis, particularly for Alzheimer's disease diagnosis. However, these methods still have limitations, such as the dependence on templates, low efficiency in runtime, and sensitivity to noise. In contrast, our proposed method overcomes these limitations and achieves superior efficiency and accuracy by utilizing a data-driven and end-to-end process based on the proposed brain region-aware module and graph contrastive learning.

\section{Methodology}
As shown in Fig. \ref{fig_1}, a two-stage model is proposed for medical brain imaging disease diagnosis, which includes a Brain Region-Aware Module(BRAM) for brain region localization and feature extraction and a Brain Network Optimization Module(BNOM) to obtain better brain network connection. In the first stage, the BRAM uses the standard AAL template \cite{AAL2020} to guide brain region localization and alignment to the standard brain space based on diffusion mechanism, automatically learning boundary and position information for brain regions. In the second stage, inspired by the idea of graph contrastive learning \cite{contrastive2020}, the BNOM utilizes the graph or brain network to eliminate individual differences between populations that are unrelated to diseases, generating stable and reliable brain networks.

% 3.1
\subsection{Brain Region-Aware Module}
\subsubsection{Brain region feature extraction and alignment}

As is well known, CNNs can extract local features and obtain global information representation through multi-layer stacking. In order to identify targets of different sizes, we need convolutional kernels of different sizes. For the human brain, the cerebral cortex is full of wrinkles, but the boundary of the brain area is not smooth. In addition, different brain regions have different sizes, so refined brain region division is a challenging problem. In order to accurately extract information from various brain regions, we use AAL templates as supervision and use registration models to achieve brain region alignment and capture the features of each brain region.

Image registration methods based on deep learning can estimate the nonrigid voxel correspondence between two images, namely moving and fixed image pairs. The registration methods generally find the optimal voxel deformation relationship between the two images by optimizing the registration field according to the following formula:

\begin{equation}
	\psi^* = \arg \min _ \psi \mathcal L_ {sim}(f, m \circ \psi) + \mathcal L_ {reg}(\psi)
\end{equation}
where $\psi^*$ is the optimal registration field for transforming a moving image into a fixed image, and $m$ is the moving image and $f$ is the fixed image. $\mathcal  L_{sim}$ is the similarity function, which is used to calculate the similarity between the deformed image and the fixed image, and $\mathcal L_{reg}$ is the regularization penalty for the registration field.

\subsubsection{The structure of proposed BRAM}

Traditional medical registration models typically consist of two networks, where the registration network receives inputs from $f$ and $m$ to generate a registration field. Then input $f$ and the registration field into the spatial transformation network to obtain the final deformation image. We propose a new unsupervised image registration method based on the diffusion model. The image registration method, by introducing the powerful image generation ability of diffusion, generates the image deformation field in the first stage from noise. RAM consists of two sub-networks: denoising dual U-Net and deformation network, the same spatial transformation network in VoxelMorph \cite{voxelmorph}.

The denoising network estimates the randomly increasing noise in the diffusion forward process, while the deformable network accepts the estimated noise input and the moving image $m$, and outputs the deformation field. We introduce time embedding into the network to enable the deformable network to learn image features under different noise levels. Since the input DTI image is 3D, the deformation function in the deformation network uses Trilinear interpolation.

The registered image warps the moving image $m$ by using the spatial transformation layer \cite{STnetwork}. For 3D image transformation, the transformation function of Trilinear interpolation is used. Obtaining deformed image $M'$ through transformation:
\begin{equation}
	M' =  M \circ\psi
\end{equation}

The final output of the first stage model is:
\begin{equation}
	M_{i}=\ M'_i \cdot\ \text{mask}({region}_i)
\end{equation}

\subsubsection{Reverse alignment process}
The reverse diffusion process is mainly based on the main structure of the DDPM\cite{20}, and the reverse process is used to estimate a conditional score function with a Diffusion network $D_\theta$. Specifically, for a moving source image $m$ and a fixed reference image $f$, the Diffusion network $D_\theta$ is trained to learn the conditional score function of the deformation between the moving image and the fixed image under the given condition $c = (m, f)$. Therefore, the target latent variable $x_t$ is sampled, and the fixed image is defined as the target, i.e., $x_0 = f$. The loss function of the Diffusion network is as follows:
\begin{equation}
	\mathcal{L}_{\mathrm{diffusion}}\left(c,x_t,t\right)=\mathbb{E}_{\epsilon,x_t,t}\left.||D_\theta\left(c,x_t,t\right)-\epsilon\right.{||}_2^2
\end{equation}

The optimization objective of the Diffusion model is:
\begin{equation}
	\min _{D_\theta,M_\psi}{\mathcal{L}_{\mathrm{diffusion}}}\left(c,x_t,t\right)+\lambda \mathcal{L}_{\mathrm{regist\ }}\left(m,f\right)
\end{equation}
where $\mathcal{L}_{\mathrm{regist\ }}$ is the registration loss, which includes two loss functions of image constraint, $\mathcal L_{\mathrm{sim}}$ and $\mathcal L_{\mathrm{smoo}}$. $M$ is the input moving images, $F$ is the input reference images. Finally, the optimization objective of this module is

\begin{equation}
	\psi=\arg\min{\mathcal {L}}_{\mathrm{sim\ }}\left(F,M\circ\psi\right)+\mu \mathcal{L}_{\mathrm{smoo}}\left(\psi\right)
\end{equation}
where ${\mathcal {L}}_ { \mathrm {sim}}$ is the loss of similarity in image structure, $\mathcal{L}_ { \mathrm {smoo}}$ is a smoothness constraint on the gradient of the deformation field \cite{enhancing2021}.

After training the Diffusion network $D_\theta$, we use the following Algorithm \ref{alg1} to generate aligned images from the original native spaces.

\begin{algorithm}[!h]
	\caption{Aligned image generation process}\label{alg1}
	\SetKwInOut{Input}{Input}\SetKwInOut{Output}{Result}
	\begin{mdframed}[backgroundcolor=bottomcolor,rightline=false,leftline=false,topline=false,bottomline=false,innerleftmargin=10pt,innerrightmargin=10pt,userdefinedwidth=230,innerbottommargin=5pt,innertopmargin=5pt]
		
	\Input {Conditional images $c=(m, f)$, where $m$ denotes the moving image in native space and $f$ denotes the fixed image in MNI space.}
	\textsc{Set } $ T \in (0, T_{train}) $ \;
	\textsc{Sample } $ x_T = \sqrt{\bar \alpha_T} m + \sqrt{1-\bar \alpha_T} \epsilon \text{, where } \bar \alpha_t = \prod_i^t \alpha_i \text{, and }\epsilon  \sim \mathcal{N}(0, I)$ \;
	\For{ $t=T, T-1, \cdots, 1$ }  {
		$ z \sim \mathcal{N}(0, I) $ \;
		$ x_{t-1} = \frac {1}{\sqrt{\alpha_t}}(x_t - \frac{\beta_t}{\sqrt{1-\bar \alpha_t}}D_\theta(c, x_t, t))+\sigma_t z $ \;
	}
	\textbf{return } $ x_0 $ \;
	\Output{Deformed image to align.}
	\end{mdframed}
\end{algorithm}

% 3.2
\subsection{Brain Network Optimization Module}

% htbp
\begin{figure*}[!t]
	\centering
	\includegraphics[width=\textwidth]{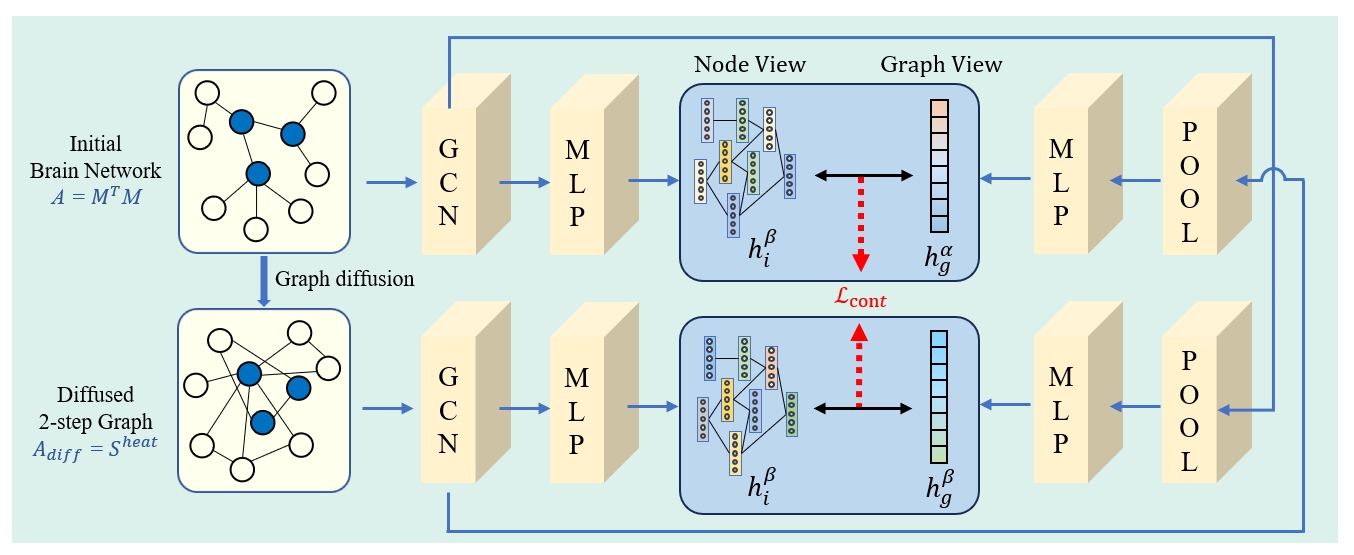}
	\caption{The detailed architecture of Brain Network Optimization Module. The network takes two graphs as input, one is the original adjacency matrix $A$ and another is the diffusion matrix $A_{\text{diff}}$. The diffusion matrix is an augmentation of the original matrix, providing an additional structural view. This model consists of two parallel GCNs \cite{gcn2017}, where pooling and MLP share weights to learn node representations in brain regions. The learned features are then fed to a graph pooling layer to obtain the final graph embedding representation.}
	\label{fig_2}
\end{figure*}

Mining brain network representations using single-modal images is challenging. To address this issue, this section introduces a strategy of graph contrastive learning. Using graph diffusion technology, the extracted original brain network graph and the diffused graph based on hot kernel features are simultaneously input into the graph neural network representation reconstruction module, And based on the similarity of node embedding and graph embedding features, the model is guided to learn the common connections between similar samples and the differences between different categories, improving the performance of the initial single mode constructed brain network.

Reconstructing the embedding features of brain network nodes involves the following three steps. 1) Sample the original brain network separately and construct a set of subgraphs; 2) Aggregating node features of subgraphs and diffused graphs through two layers of GCN; 3) Calculate the similarity of node embedding and graph embedding by contrastive loss function. The specific design of this module can be seen in Fig. \ref{fig_2}.

The diffusion map is generated by the thermonuclear features, and the formula is as follows:
\begin{equation}
	S^{\mathrm{heat\ }}=\exp{\left(tAD^{-1}-t\right)}
\end{equation}
where $A$ represents the adjacency matrix, $D$ represents the corresponding degree matrix, and $t$ is the diffusion time \cite{diffusion2019}.
The diffusion process on the graph can be seen as introducing local information within the neighborhood to the node. The subsampling method involves randomly sampling nodes and their edges from a graph, and selecting precise nodes and edges from another graph, with each graph serving as a different view.

The loss function designed for contrastive learning is as follows:
\begin{equation}
	\mathcal{L}_{\text {cont}}=\max _{\theta, \omega, \phi, \psi} \frac{1}{|\mathcal{G}|} \sum_{g \in G} \sum_{i=1}^{|g|}\left [\text{MI} (h_i^\alpha, h_g^\beta)+ \text{MI}(h_i^\beta, h_g^\alpha)\right]
\end{equation}
where \text{MI} represents the mutual information function.

Graph contrastive learning of brain networks involves the following processes: 1) Extract graph embedding features from node features through the readout function; 2) Calculate the dot product between node embedding and graph embedding on another view; 3) Update model losses based on similarity, calculate loss gradients, and update network weights.

In order to keep the symmetry of the generated brain network matrix, we conduct matrix inner product operation on the extracted embedding, and limit it to a range of 0 to 1 through the activation function \cite{structuralbrain}. The reconstruction loss of edges adopts mean square error:
\begin{equation}
	\mathcal{L}_{\mathrm{ref}}\left(\mathcal{G}\right)=-\frac{1}{90\times90} \sum _{i,j}(A_p[i,j]-A_{\mathrm{ref}}[i,j])^2
\end{equation}
where $A_{p}[i, j]=\sigma\left(h_{i}^{\top} h_{j}\right)$.

% 3.3
\subsection{Classifier Design and Model Training}
The proposed DGCL ultimately guides brain network generation through disease classification task. The classifier knowledge contains a graphical representation of rich pathological information through multi-layer perceptron aggregation, and is connected with the Softmax function to output the prediction probability of each category of diseases. This module consists of three layers of MLP (Multilayer Perceptron), two layers of ReLU activation function and the Softmax output layer. The Softmax input contains four neurons, mapping the results to the probabilities of four disease categories. The Dropout strategy is used to prevent the model from overfitting \cite{dropout}.

Multi-class cross entropy function is adopted for multi-classification loss \cite{dementia2019}. In our study, the number of classes $C$ is 4.
\begin{equation}
	\mathcal{L}_{\mathrm{cls}}\left(p\right)=-\sum_{i=0}^{C-1}{y_i\log (p_i)}+({1-y}_i)\log (1-p_i)
\end{equation}

At last, the total loss function of DGCL is as follows:

\begin{equation}
	{\mathcal{L}_{total}=\mathcal{L}_{\mathrm{cont}}+\mathcal{L}}_{\mathrm{cls}}\left(p\right)+{\alpha\mathcal{L}}_{\mathrm{ref}(\mathcal{G})}
\end{equation}

The overall pipeline of the training process can be summarized in Algorithm \ref{alg2}.

\begin{algorithm*}[!ht]	
	\caption{The end-to-end brain network construction pipeline}
	\label{alg2}
	\begin{mdframed}[backgroundcolor=bottomcolor,rightline=false,leftline=false,topline=false,bottomline=false,innerleftmargin=15pt,innerrightmargin=10pt,userdefinedwidth=490,innerbottommargin=5pt,innertopmargin=5pt]
	
	\SetKwInOut{Input}{Input} \SetKwInOut{Output}{Result}
	\Input{$X$: training dataset of original brain images, and each image is a moving image, and $x_{mni}$ represents the fixed image; $A_{ref}$ : the empirical connection matrix of brain network;
		$Y$: the labels of corresponding subjects;  $T$: the maximum timestamp; $N$ : the number of training epochs.}
	\BlankLine
	{Randomly initialize the parameters of Dual U-Net model}\;
	\For{$s\leftarrow 1$ \KwTo $N$} {
		\For {$x \in X$} {
			Train the Denoising Dual U-Net\;
			\textbf{Repeat}  \;
			\hspace{0.5cm} $x \sim q(x_0)$ \;
			\hspace{0.5cm} $t \sim \text{Uniform} (\{1,\dots, T\})$ \;
			\hspace{0.5cm} $\epsilon \sim \mathcal{N}(0, \mathbf{I})$ \;
			\hspace{0.5cm} Take gradient descent step on  $\nabla_{\theta}\left\|\epsilon-{D}_{\theta}\left(\sqrt{\bar{\alpha}_{t}} \mathbf{x}_{0}+\sqrt{1-\bar{\alpha}_{t}} \epsilon, t\right)\right\|^{2}$ \;
			\textbf{until converged} \;
			
			$\epsilon^f = D_\theta(c, f, 0) $ \;
			\For {$ \eta \in [0, 1]$} {
				$\epsilon _\eta ^f \gets \eta \cdot \epsilon^f $ \;
				$\phi_n \gets M_\psi(m, \epsilon_\eta ^f) $ \;
			}
			\textbf{Obtain: } The aligned brain regions $ m(\phi_n)$ \;
		}
		\BlankLine
		\textbf{Calculate: } \;
		The feature of each brain region: $M_{region_i} = M_i \cdot \text{mask}(region_i)$ \;
		The initial brain network: ${A}= M^T M$ \;
		$\mathcal{L}_{\mathrm{BRAM}} = \mathcal{L}_{\mathrm{diffusion}} \left(c,x_t,t\right)+\lambda \mathcal{L}_{\mathrm{regist\ }}\left(m,f\right)$ \;
		
		\For {$x \in X$} {
			$A_\mathrm{diff} = S^{\mathrm{heat\ }}=\exp{\left(tA D^{-1}-t\right)}$ \;
			$(h^{\alpha}_g, h^{\beta}_g) = \mathrm{MLP(GCN((}A, A_\mathrm{diff}))$ \;
			$(h^{\alpha}_i, h^{\beta}_i) = \mathrm{MLP(POOL(GCN((}A, A_\mathrm{diff})))$
		}
		
		$A_p = H_i^T H_i$ \;
		
		$\mathcal{L}_{\mathrm{cont}}=\max _{\theta, \omega, \phi, \psi} \frac{1}{|\mathcal{G}|} \sum_{g \in G} \sum_{i=1}^{|g|}\left [\text{MI} (h_i^\alpha, h_g^\beta)+ \text{MI}(h_i^\beta, h_g^\alpha)\right]$ \;
		
		$\mathcal{L}_{\mathrm{ref}}\left(\mathcal{G}\right)=-\frac{1}{90\times90} \sum _{i,j}(A_p[i,j]-A_{\mathrm{ref}}[i,j])^2$  \;
		
		$\mathcal{L}_{\mathrm{cls}}\left(p\right)=-\sum_{i=0}^{C-1}{y_i\log (p_i)}+({1-y}_i)\log (1-p_i)$ \;
		
		\textbf{Calculate:} total loss function $\mathcal{L}_{total}=\sum^N_{t=1} {\mathcal{L}_{\mathrm{cont}}+\mathcal{L}}_{\mathrm{cls}}\left(p\right)+{\alpha\mathcal{L}}_{\mathrm{ref}}(\mathcal{G}) $ \;
		{Back propagation and update the parameters} \;
	}
	\Output{The generated brain network $A_p$.}
	\end{mdframed}
\end{algorithm*}

\section{Experiments}
% 4.1
\subsection{Datasets and Preprocessing}
This study utilizes DTI from the Alzheimer's Disease Neuroimaging Initiative (ADNI) \cite{ADNI2005}, an open-source and public dataset to validate the proposed framework. A total of 349 subjects' data are collected, including normal control group (NC), early mild cognitive impairment (EMCI), late mild cognitive impairment (LMCI), and AD. TABLE \ref{table1} provides detailed information about the sample size, gender, and age of all subjects. The PANDA toolbox \cite{panda2013} is used for preprocessing the raw DTI data to obtain the reference structural brain network matrix. The process involves converting the initial DICOM format of the data to NIFTI format, skull stripping, fiber bundle resampling, and head motion correction. Then we calculate the fractional anisotropy (FA) coefficients by fitting the tensor model using the least squares method, and output the DTI data. After resampling, all subjects have $91 \times 109 \times 91$ voxels in DTI, with a voxel size of $2mm \times 2mm \times 2mm$. After that, we register images to the individual brain space, and construct the empirical structural connectivity $\hat{A}$ based on the deterministic fiber tracking method by setting tracking conditions, network nodes, and tracking stopping conditions. The brain is divided into 90 ROIs based on the AAL atlas \cite{AAL2020}, with each ROI defined as a node in the brain network. Finally, the structural connectivity of the brain network is determined by fiber tracking between different ROIs. Specifically, the stopping criteria for fiber tracking are defined as follows: (1) the crossing angle between two consecutive directions is greater than 45 degrees, and (2) the anisotropy score value is not within the range of [0.2, 1.0].

\begin{table}[!ht]
	\caption{Subjects’ information in this study.\label{table1}}
	\centering
	\begin{tabular}{ccccc}
		\toprule
		Group & NC(87) & EMCI(135) & LMCI(63) & AD(64) \\
		\midrule
		Male/Female & 46M/41F  & 83M/72F  & 26M/37F  & 35M/29F \\
		Age(mean$\pm$std) & 74.3$\pm$5.5 & 74.9$\pm$5.8 & 75.8$\pm$6.1 & 75.6$\pm$5.4 \\
		\bottomrule
	\end{tabular}
\end{table}

% 4.2
\subsection{Experiment Settings}
The model is trained and tested using the PyTorch platform, with an NVIDIA RTX 4000 GPU with 20GB memory. During the training process, the optimizer is set to Adam \cite{adam2014}, with an initial learning rate of 0.0001, which exponentially decays with the number of training iterations. The number of epochs is set to 300, and the batch size is set to 16. Five-fold cross-validation is used to evaluate the performance of the model. All subjects are randomly divided into five equally sized subsets. One subset is treated as the test set, and the union of the other four subsets is treated as the training set. This process is repeated five times to eliminate bias. The evaluation metrics for the classification performance in this study are accuracy (ACC), sensitivity (SEN), specificity (SPE), and the area under the ROC curve (AUC).

The learnable parameter in BRAM is initialized according to the Xavier scheme. The structure of BNOM is described in detail in section 3.2. We use two GCNs to progressively extract high-order topological features of brain regions, with a brain region feature vector dimension of $d=128$. The learnable parameter is initialized with the identity matrix. The Discriminator consists of two FC layers, with a LeakyReLU activation function in the hidden layers and no activation function in the output layer. The classification feature $h \in \mathbb{R}^{90 \times 128}$ is inputted into the FC layer for disease prediction, with 128 neurons in the input layer and 4 neurons in the output layer.

% 4.3
\subsection{Ablation Study and Analysis}
In this study, we conduct experiments to investigate the effect of two hyperparameters, namely, $\lambda$ and $\mu$, in the loss function of the BRAM. $\lambda$ represents the loss coefficient of the alignment network, while $\mu$ represents the constraint coefficient of the smoothness of the deformation field gradient. We also analyze the effect of the $\alpha$ hyperparameter, which is the coefficient of the reference brain network, on the classification performance during the joint training phase. $\lambda$ is set to 2.0, the optimal value according to the original paper, representing the ratio between the diffusion model and the registration network weights. Then we adjust the parameter $\mu$ in the following experiments. In the brain network optimization phase, we evaluate the ACC and AUC values of the binary classification task (AD vs. NC) by changing $\alpha$  from 0.1 to 0.9 while keeping other parameters constant.

\begin{table}[!ht]
	\caption{The quality of alignment and registration results under different hyperparameters. \label{table2}}
	\centering
	\begin{tabular}{ccccc}
		\toprule
		Method & hyperparameters & learning rate & PSNR & Dice  \\
		\midrule
		ANTs & - & - & - & 0.694  \\
		BRAM & $\lambda=2.0, \mu=0.3$ & $1 \times 10^{-4}$ & 27.3 & 0.671  \\
		BRAM & $\lambda=2.0, \mu=0.5$ & $1 \times 10^{-4}$ & 27.9 & 0.735  \\
		BRAM & $\lambda=2.0, \mu=0.8$ & $1 \times 10^{-4}$ & 28.7 & 0.798  \\
		BRAM & $\lambda=2.0, \mu=1.0$ & $1 \times 10^{-4}$ & 28.3 & 0.786  \\
		\bottomrule
	\end{tabular}
\end{table}

From the experimental results in TABLE \ref{table2} above, it can be seen that there is little difference between different methods, but the BRAM module using the diffusion mechanism can get the best effect, in which the Dice coefficient under the hyperparameters $\lambda$=2.0 and $\mu$=0.8 can reach the optimal 0.798, far exceeding the effect of the ANTs software.

\begin{table}[!ht]
	\caption{The quality of brain networks constructed under different hyperparameters. \label{table3}}
	\centering
	\begin{tabular}{cccccc}
		\toprule
		$\alpha$ & Epoch & ACC & AUC & Quality & Identifiability \\
		\midrule
		0.1 & 42 & 0.953 & 0.961 & bad & strong\\
		0.3 & 53 & 0.964 & 0.974 & bad & strong\\
		0.5 & 69 & 0.963 & 0.971 & good & weak\\
		0.7 & 147 & 0.971 & 0.974 & good & strong\\
		0.9 & 231 & 0.965 & 0.968 & superior & strong\\
		\bottomrule
	\end{tabular}
\end{table}

The results show that there is no significant difference between the different methods in the brain area localization phase. However, using the diffusion mechanism yield the best results, with a Dice coefficient very close to the optimal value of 0.802 in the original paper, with $\lambda$ set to 2.0 and $\mu$ set to 0.8. To measure the difference between the two networks, we choose from several loss functions, including cross-entropy, absolute loss, and square loss (including mean and sum modes). We find that choosing the square loss function yielded better results, with the sum mode being equivalent to punishing every edge, resulting in the slowest convergence rate. The mean mode could converge in 40 epochs, after which they were unable to learn the features of the brain network.

% 图3
%\begin{figure}[!ht]
%	\centering
%	\includegraphics[width=3.5in]{region70.png}
%	\caption{The diffusion model achieves the effect of brain region localization in the first stage, taking the 70 brain region as an example.}
%	\label{fig_3}
%\end{figure}

\begin{figure}[!ht]
	\centering
	\subfloat[$\alpha$=0.3]{\includegraphics[width=3.5in]{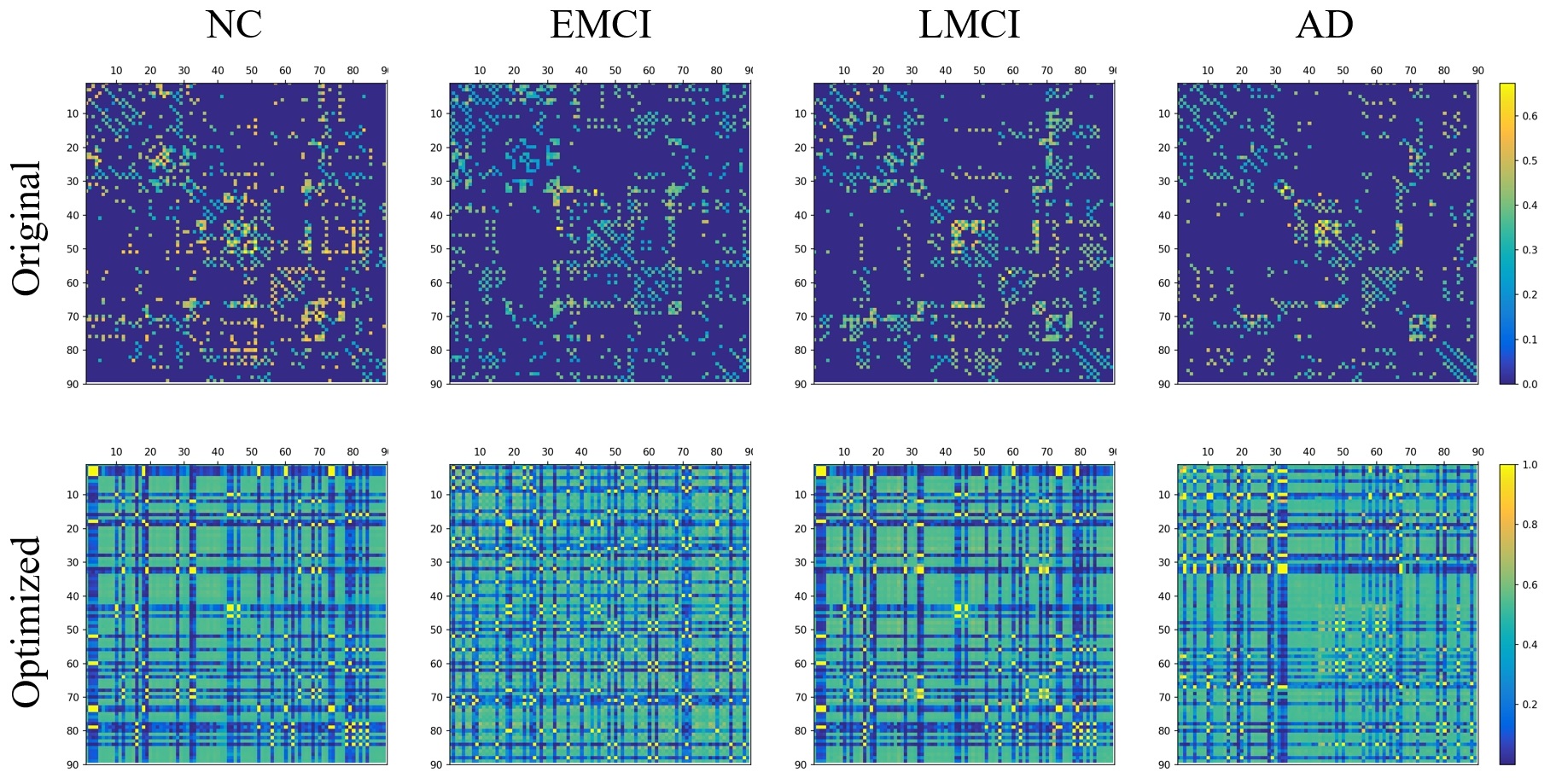}}
	\vfil
	\subfloat[$\alpha$=0.9]{\includegraphics[width=3.5in]{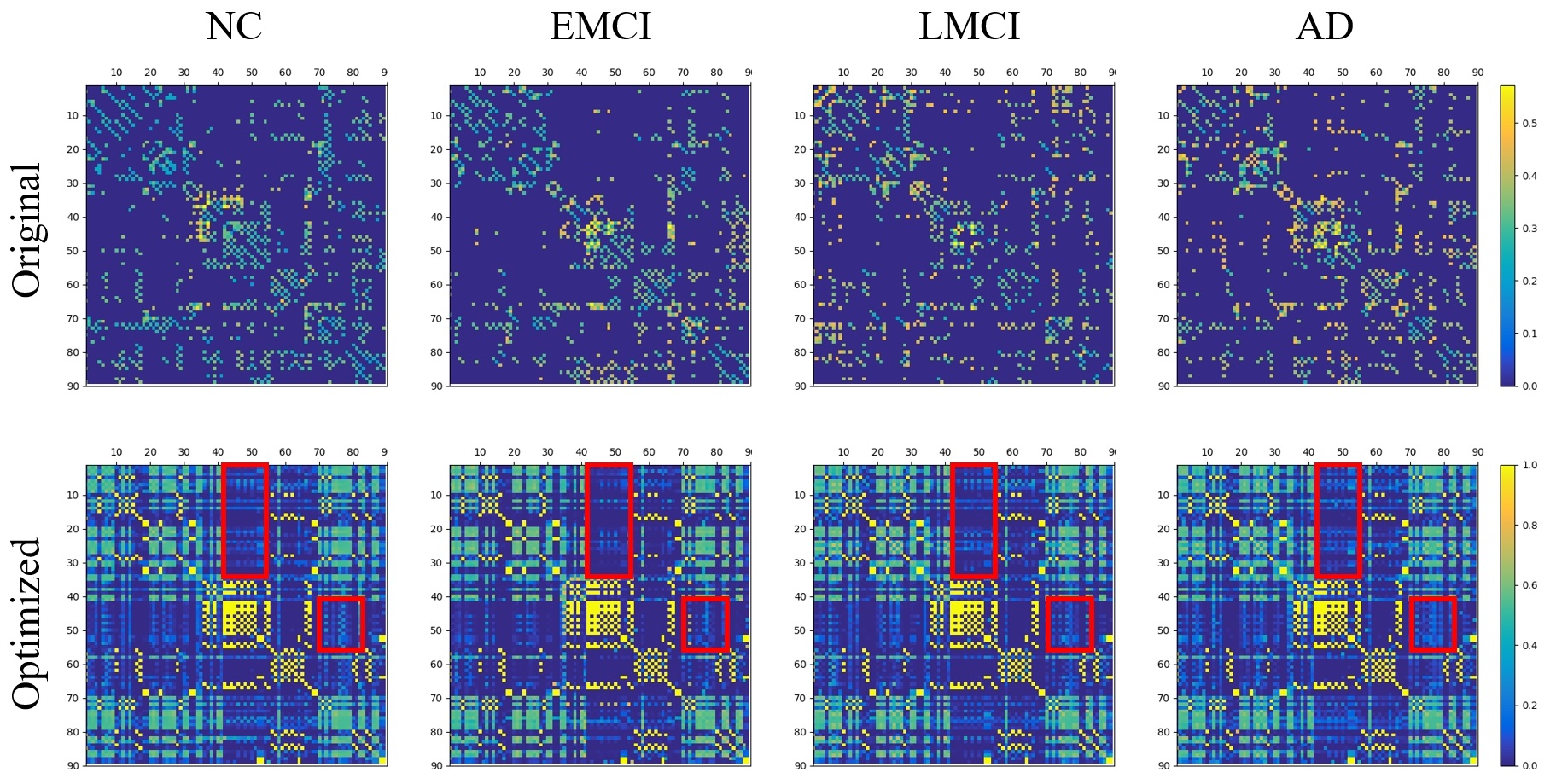}}
	\caption{Visualization of brain network matrix under different hyperparameters. The first row represents the original brain network matrix, and the second row represents the optimized brain network.}
	\label{fig_4}
\end{figure}

The hyperparameter $\alpha$ mainly affects the brain generation quality. The results are shown in TABLE \ref{table3}. The two columns of Quality and Identifiability are judged by experienced experts in the field of brain science and cognitive neuroscience. When $\alpha$  is set to 0.9, the model converged after 231 epochs of training. At this point, the optimized brain networks for different classes have good discriminative power, and exhibit similar patterns to the reference brain network as shown in Fig. \ref{fig_4}. Moreover, the brain networks constructed by the proposed DGCL have high consistency on the same group of samples.

We propose the BRAM for brain atlas based on the diffusion model and the BNOM based on graph contrastive learning. We conduct ablation experiments on these two modules, namely BRAM implemented using the DDPM model and BNOM implemented using the GCN model. When the BNOM is not used, two fully connected layers (MLP) are employed for multiclass classification. The results (TABLE \ref{table4}) show that the brain area localization module has a relatively small effect on performance, while the BNOM for brain network has a more significant impact on performance improvement. Using an unaligned brain area without location information for subsequent classification results in poorer performance compared to the results obtained using the PANDA software. The initial brain network obtained using the brain area localization module is better than the experience-based brain network obtained from the software, even without optimization. The ablation experiments demonstrate the effectiveness of these two modules.

\begin{table}[!t]
	\caption{Results of ablation study on each block proposed. \label{table4}}
	\centering
	\begin{tabular}{ccccc}
		\toprule
		Method & AUC & ACC & SEN & SPE \\
		\midrule
		without BRAM & 0.9536 & 0.9138 & 0.9625 & 0.8353 \\
		without BNOM  & 0.8914 & 0.8839 & 0.9286 & 0.7867 \\
		BRAM+BNOM & 0.9781 & 0.9327 & 0.9550 & 0.8625 \\
		\bottomrule
	\end{tabular}
\end{table}

\begin{figure}[!ht]
	\centering
	\includegraphics[width=3.5in]{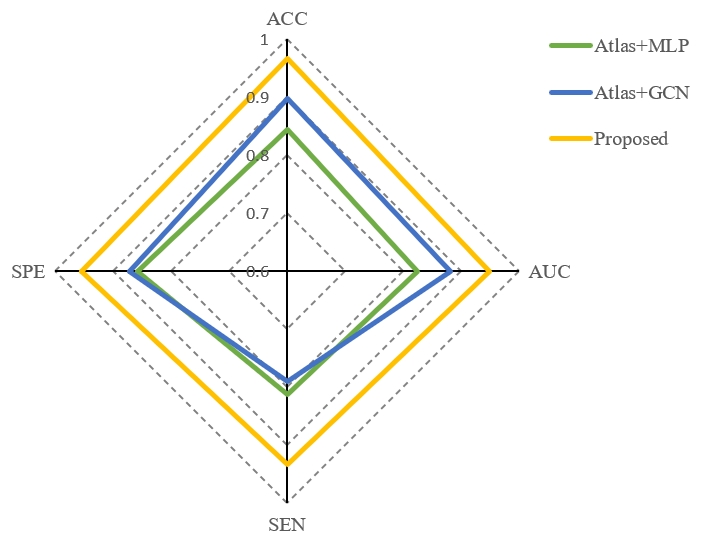}
	\caption{Comparisons of AD classification performance on different metrics between proposed method with other two basic methods.}
	\label{fig_5}
\end{figure}

\begin{figure}[!ht]
	\centering
	\includegraphics[width=3.5in]{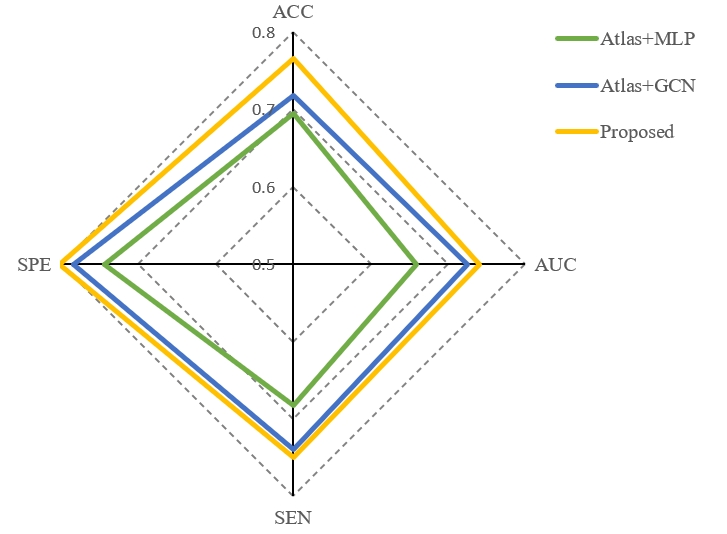}
	\caption{Comparisons of ASD classification performance on different metrics between proposed method with other two basic methods. The classification performance of the proposed model on two datasets is comprehensively better than that of the comparison method in all indicators.}
	\label{fig_6}
\end{figure}

\section{Discussion}
% 5.1
\subsection{Comparisons with Baseline}
The brain network constructed by the PANDA software plus the dual-layer GCN/four-layer MLP was used as a baseline for experimental comparison. A four-class performance comparison was designed, with each metric result representing the average value of the one vs. rest binary classification corresponding indicators for the four classes. Compared with directly using a four-layer fully connected layer for classification, the proposed method performed significantly better in all four indicators, outperforming other methods. The ACC indicator increased by 14.47\%, the AUC indicator increased by 11.66\%, the sensitivity increased by 9.65\%, and the specificity increased by 17.60\%. Although the performance slightly improved by replacing the first two fully connected layers with GCN, it still did not reach the performance achieved by the initial brain network constructed based on the region localization method proposed in this study. The results suggest that introducing the region localization module is advantageous for improving the model's disease detection performance.

\begin{table*}[!ht]
	\caption{Results of 5-fold cross validation on each category of AD,  and values are reported as mean $\pm$ standard deviation. \label{table5}}
	\begin{adjustbox}{width=2\columnwidth,center}
		\begin{tabular}{ccccccccc}
			\toprule
			& \multicolumn{4}{c}{Validation Set}                                             & \multicolumn{4}{c}{Test Set}                                              \\
			\midrule
			categary & AUC & ACC & SEN & SPE & AUC & ACC & SEN & SPE\\
			\midrule
			NC & 0.9583$\pm$0.0563 & 0.8957$\pm$0.0884 & 0.9850$\pm$0.0200 & 0.8641$\pm$0.1201 & 0.8641$\pm$0.0381  & 0.8489$\pm$0.0311 & 0.8600$\pm$0.0490 & 0.8451$\pm$0.0362 \\
			EMCI & 0.9648$\pm$0.0516 & 0.9126$\pm$0.1004 & 0.9400$\pm$0.0344 & 0.9019$\pm$0.1331 & 0.8699$\pm$0.0213 & 0.8279$\pm$0.0535 & 0.8309$\pm$0.0961 & 0.8259$\pm$0.0839 \\
			LMCI & 0.9670$\pm$0.0268 & 0.8722$\pm$0.0816 & 0.9886$\pm$0.0229 & 0.8377$\pm$0.1048 & 0.9308$\pm$0.0506 & 0.8912$\pm$0.0751 & 0.9111$\pm$0.0831 & 0.8855$\pm$0.0933 \\
			AD & 0.9706$\pm$0.0208 & 0.9036$\pm$0.0663 & 0.9719$\pm$0.0354 & 0.8835$\pm$0.0911 & 0.8384$\pm$0.0931 & 0.8016$\pm$0.0625 & 0.8000$\pm$0.2475 & 0.8028$\pm$0.1381 \\
			\bottomrule
		\end{tabular}
	\end{adjustbox}
\end{table*}

\begin{figure}[!h]
	\centering
	\includegraphics[width=3.5in]{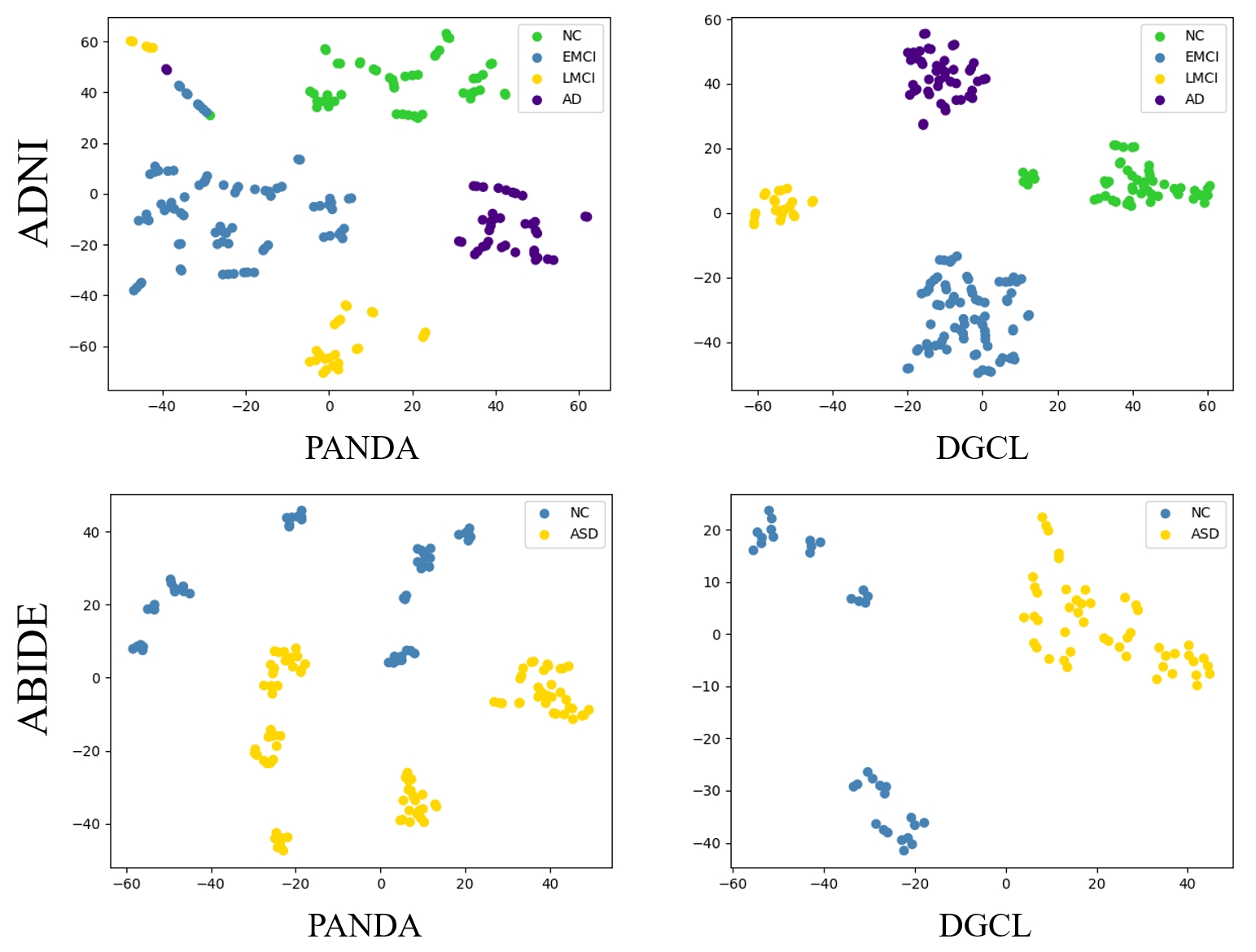}
	\caption{Visualization of the features of different disease categories of the comparative and proposed methods under the t-SNE tool. The above two figures show the dimensionality reduction visualization results of brain networks constructed by DGCL and PANDA on the dataset ADNI. The bellow figures show the feature visualization on the dataset ABIDE under the same conditions. Compared to the brain network features generated by PANDA, the proposed DGCL can generate planes that are easier to segment.}
	\label{fig_7}
\end{figure}

Fig. \ref{fig_5} and Fig.\ref{fig_6} compare the performance indicators of different methods on AD and ASD datasets, respectively. Meanwhile, we further studied our model's classification performance through t-SNE analysis. Fig. \ref{fig_7} compares the visualization results of the brain network embedding features of different methods, showing the two-dimensional projection of the learned brain network features of three methods.

Compared with directly adopting the brain network constructed by the PANDA software, our model can generate more easily segmented planes, indicating that the features obtained by our method are more discriminative than those of PANDA. The visualization results also intuitively explain why our model performs better than other models in disease classification tasks as show in Fig. \ref{fig_7}.

To evaluate the quality of generating brain networks, multiple comprehensive indicators are used to compare the differences between the proposed method and the software. Among them, ICC is the most commonly used measure in assessing the reliability \cite{hu2023phipipe}. TABLE \ref{table6} shows that the proposed model has better performance on each dimension metric.

\begin{table}[!h]
	\caption{The comprehensive comparison of brain network construction using PANDA and DGCL.}
	\label{table6}
	\centering
	\begin{adjustbox}{width=\columnwidth,center}
		\begin{tabular}{ccccc}
			\toprule
			\textbf{Datasets}  & \textbf{Methods}   & \textbf{Time(s)}   & \textbf{Reliability}   & \textbf{Parameter Dependency}  \\
			\midrule
			\multirow{2}{*}{ADNI}  & PANDA & $7.32 \times 10^2$   & 0.73   & technician-dependent    \\
			& DGCL  & $\mathbf{2.23 \times 10^0}$   & \textbf{0.79}   & \textbf{highly-automated}   \\
			\midrule
			\multirow{2}{*}{ABIDE}  & PANDA & $7.25 \times 10^2$ & 0.64  & technician-dependent  \\
			& DGCL & $\mathbf{2.19 \times 10^0}$    & \textbf{0.71}      & \textbf{highly-automated}    \\
			\bottomrule
		\end{tabular}
	\end{adjustbox}
\end{table}

% 4.4.2 => 5.2
\subsection{The optimized Brain Networks}
The brain network optimization module learns the differences in brain networks between different disease categories through graph comparison, enabling the classifier to better distinguish between different disease categories. To verify the expression ability of AD-related features in the optimized brain network, a classification experiment is designed to compare the predictive performance of the reference brain network and the optimized brain network at different stages of AD. This experiment achieves four-class classification for different diseases, using five-fold cross-validation, and calculated the corresponding classification indicators using one-vs-rest for each class. The performance of this model on the test set is shown in TABLE \ref{table5}. The best accuracy rates in the four binary classification tasks were $84.89\%$, $82.79\%$, $89.12\%$, and $80.16\%$, respectively.

\begin{figure}[!thbp]
	\centering
	\subfloat[The increased connection from NC to EMCI compared to PANDA.]{\includegraphics[width=3.5in]{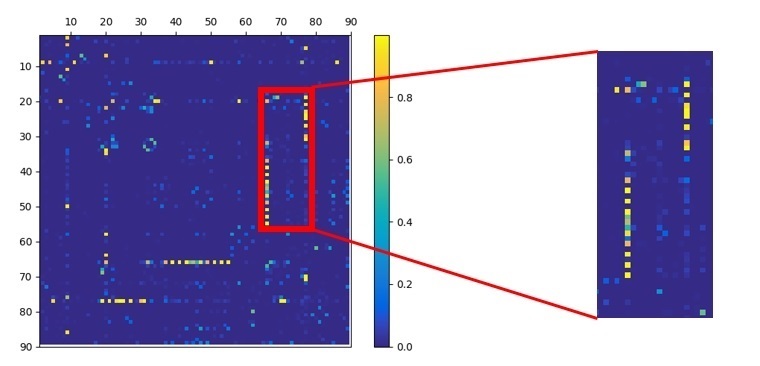}}
	\vfil
	\subfloat[The increased connection from NC to LMCI  compared to PANDA.]{\includegraphics[width=3.5in]{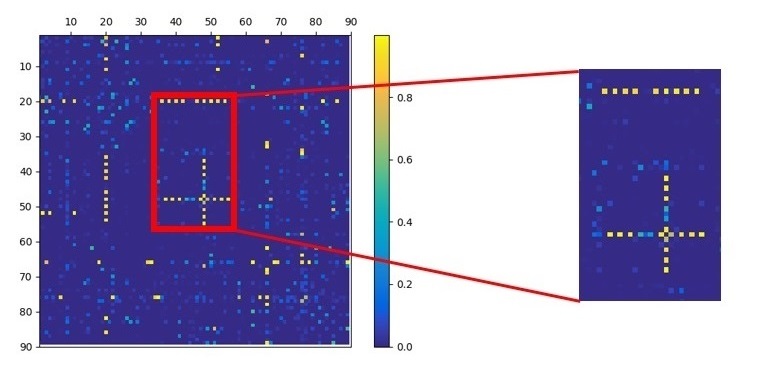}}
	\vfil
	\subfloat[The increased connection from NC to AD  compared to PANDA.]{\includegraphics[width=3.5in]{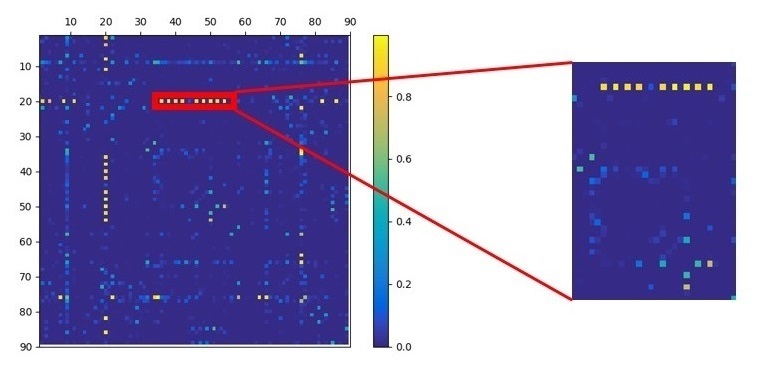}}
	\caption{The difference of brain network connection. Figures (a) - (c) show the changes in brain connectivity from NC to EMCI, LMCI, and AD, respectively.}
	\label{fig_8}
\end{figure}

\begin{figure*}[!t]
	\centering
	\includegraphics[width=0.95 \textwidth]{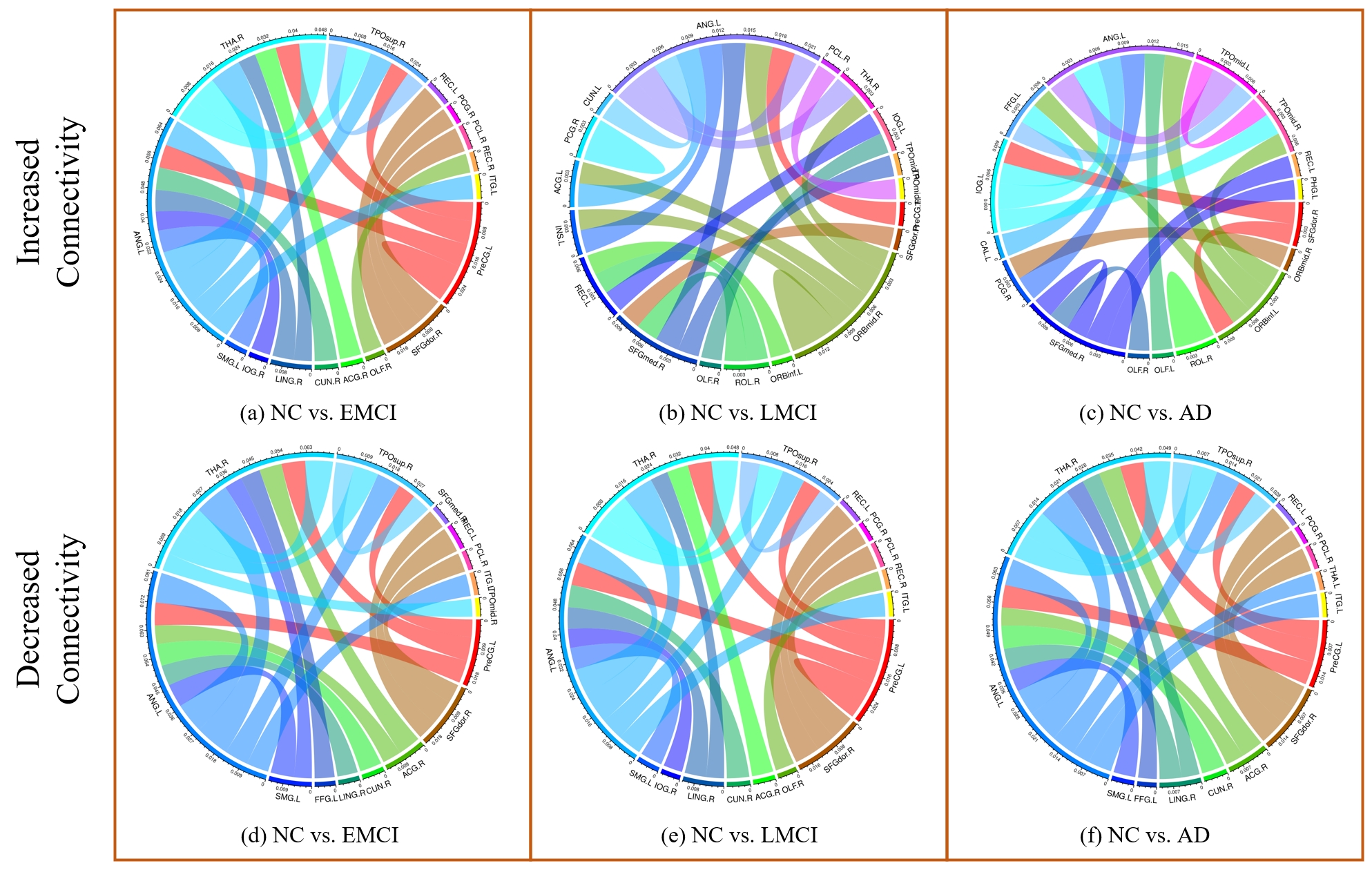}
	\caption{(a)-(c) represents the brain connections with increased connectivity, and (d)-(f) represents the brain connections with decreased connectivity. There are significant differences in the brain regions with increased abnormal connections from EMCI to LMCI and AD stages. For the decreased connections, all three stages of the disease are concentrated in the connections between the frontal lobe and other brain regions.}
	\label{fig_9}
\end{figure*}

To quantitatively analyze the changes in brain connectivity features in different cognitive disease stages, based on the output structural brain connectivity matrix of the model, we study the changes in the structural connectivity of the brain network (brain network weighted adjacency matrix) in different populations. The visualization is shown in Fig. \ref{fig_8}, which shows the brain connectivity differences between the NC vs. EMCI, NC vs. LMCI, and NC vs. AD groups, respectively. For each binary classification task, we represented the connectivity matrix of each group, subtracted the connectivity matrix of the NC group from the connectivity matrix of the other groups, and obtained the changes in brain network connectivity.

In order to intuitively understand the differences in brain network connectivity between subjects in different disease stages and normal populations, we did not threshold the results and selected the top 2\% of brain connections. From (a) to (c), it can be seen that with the deterioration of the disease, the increased connectivity initially rises in the early stage and then decreases to a low level when it worsens to AD. This result is consistent with neuroscience research, indicating that the brain may first produce compensatory mechanisms during the MCI to AD progression and weaken in the later stages of the disease.

Fig. \ref{fig_9} shows the visualization results of important brain connections in different stages of Alzheimer's disease. Figures (a)-(c) represent the increased changes in brain connections, and (d)-(f) represent the decreased changes in brain connections between different stages of the disease (NC, EMCI, LMCI, or AD). These abnormal brain connections can reflect the damage to important brain connections during the development of Alzheimer's disease. The brain regions involved in the abnormal increase of connections in EMCI and LMCI include the cingulate gyrus and occipital lobe, while in the AD stage, there is also an increase in connectivity between the frontal and temporal lobes, with the abnormal brain connections concentrated in the left cingulate gyrus and left occipital lobe. For the decreased connections, all three stages of the disease are concentrated in the connections between the frontal lobe and other brain regions. The regions with increased connectivity are consistent with most research, but the frontal lobe region is not reflected in this paper. According to \cite{fan2016}, the development of the AD brain network leads to a transformation towards a more regular network topology and a loss of diffusion in local structural connections. Compared with healthy elderly people, MCI and AD patients have reduced structural connections in the posterior cingulate gyrus of the DMN, while the decrease in the DMN posterior cingulate cortex and anterior cingulate cortex can be revealed from the above figure.

For ASD, the ROIs of the dataset can be divided into eight functional subnetworks: cerebellum and subcortical structure (CB), visual network (VN), somatic motor network (SMN), dorsal attention network (DAN), ventral attention network (VAN), edge network (LN), frontoparietal network (FPN), and default mode network (DMN).

\begin{figure}[!h]
	\centering
	\includegraphics[width=3.5in]{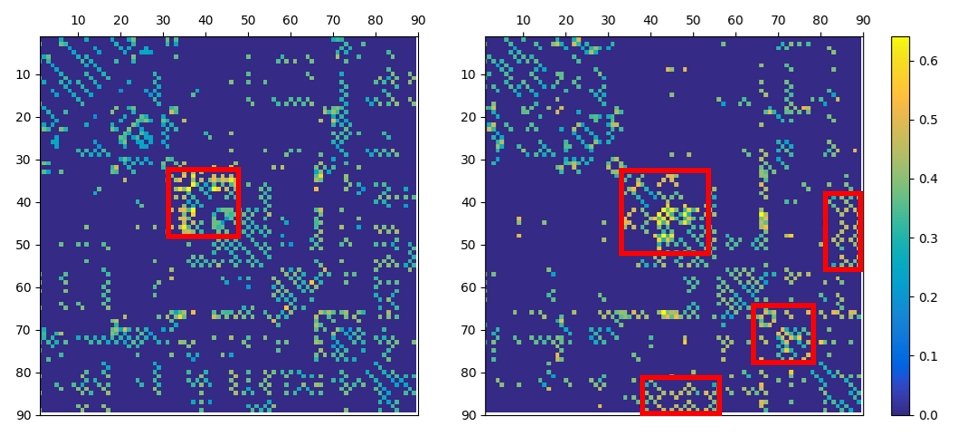}
	\caption{Subnetwork patterns of autism spectrum disorder. The red box clearly indicates the difference between HC (left) and ASD (right). In these subnetworks, there have been significant changes in brain connectivity.}
	\label{fig_10}
\end{figure}

%The dashed lines divide different subnetworks, which are CB, VN, SMN, DAN, VAN, LN, FPN, and DMN from left to right.

Fig. \ref{fig_10} shows the differences between the proposed method and the brain template (left), as well as the average matrix of the autism population with reference template (right). These matrices are obtained by averaging weight scores on correctly classified test data. Compared to the brain template, our model's learning score clearly highlights the important sub-networks for predicting ASD. Specifically, the attention score in the SMN region is high, indicating that connectivity plays a crucial role in ASD prediction in SMN.

According to relevant literature on autism, the reduction of FC in SMN reflects changes in sensory and motor processing. In addition, attention scores were higher in the DMN, DAN, and FPN regions (highlighted in the red box), indicating that functional connectivity between DMN and DAN, as well as between DMN and FPN, is also crucial for predicting ASD, consistent with previous studies reporting DMN abnormalities. Therefore, the proposed model can correctly identify functional networks related to ASD, and the results are consistent with the findings of neuroscience.

% 5.3
\subsection{Detection of disease-related brain regions}
The degree centrality of a brain region node is defined as the importance of the brain region, which is calculated using the following formula:
\begin{equation}
	C_{deg}\left(v_{i}\right)= \frac {\sum_{j} A[i, j]} {\sum_{j} \mathbf{1}(A[i,j]>0)}
\end{equation}

\begin{figure}[!h]
	\centering
	\subfloat[PANDA]{\includegraphics[width=1.0in]{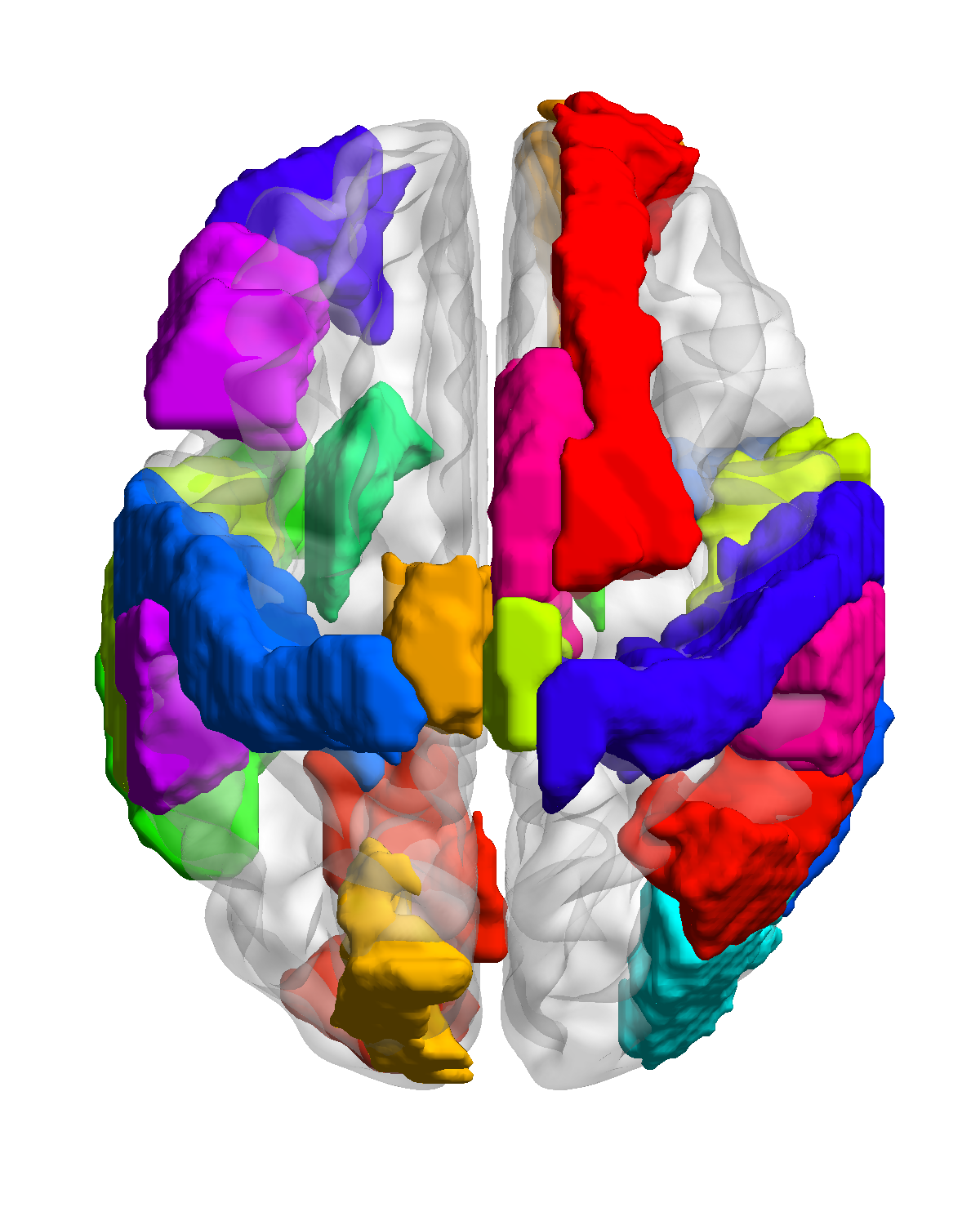}}
	\hfil
	\subfloat[DGCL]{\includegraphics[width=1.0in]{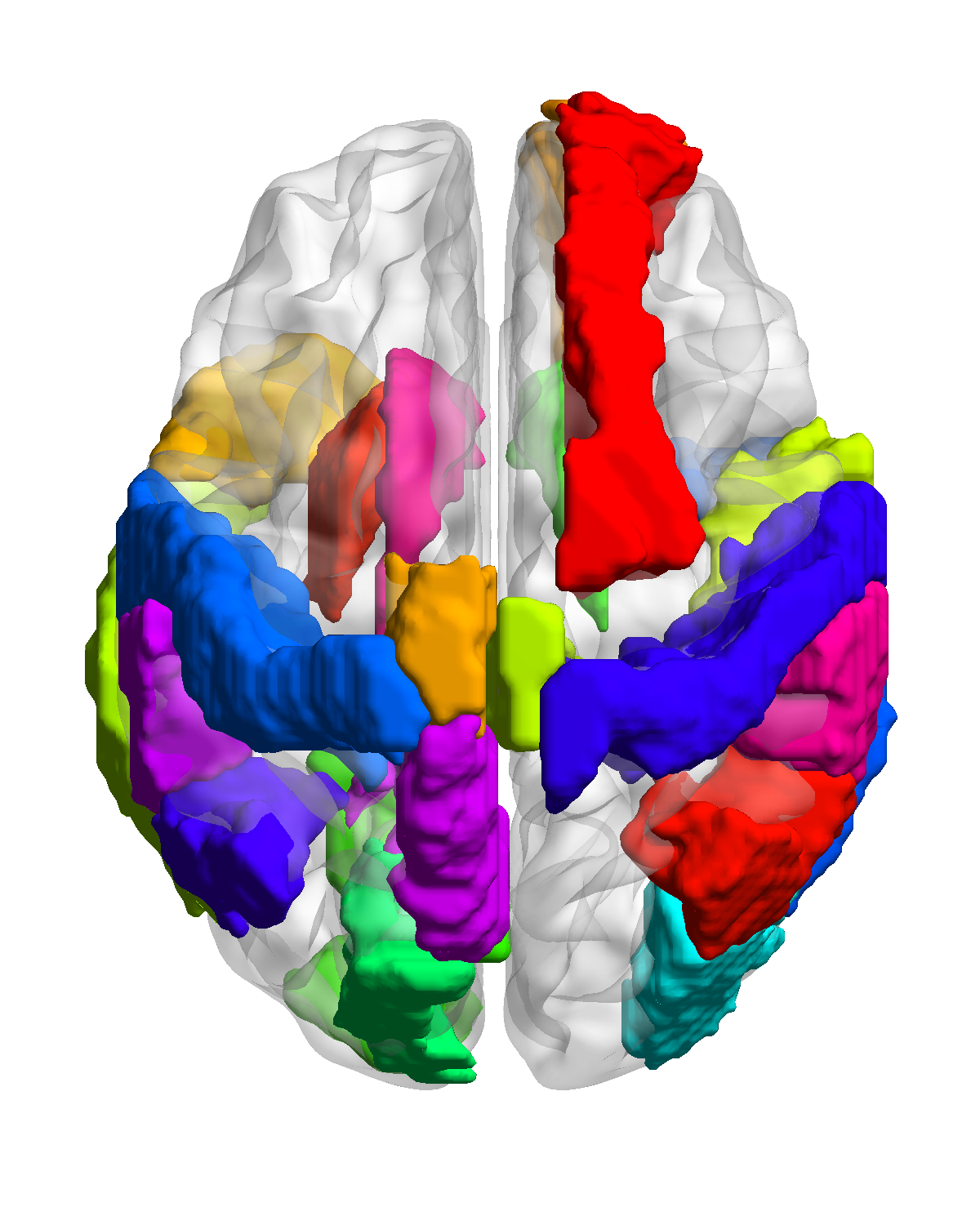}}
	\hfil
	\subfloat[Common]{\includegraphics[width=1.0in]{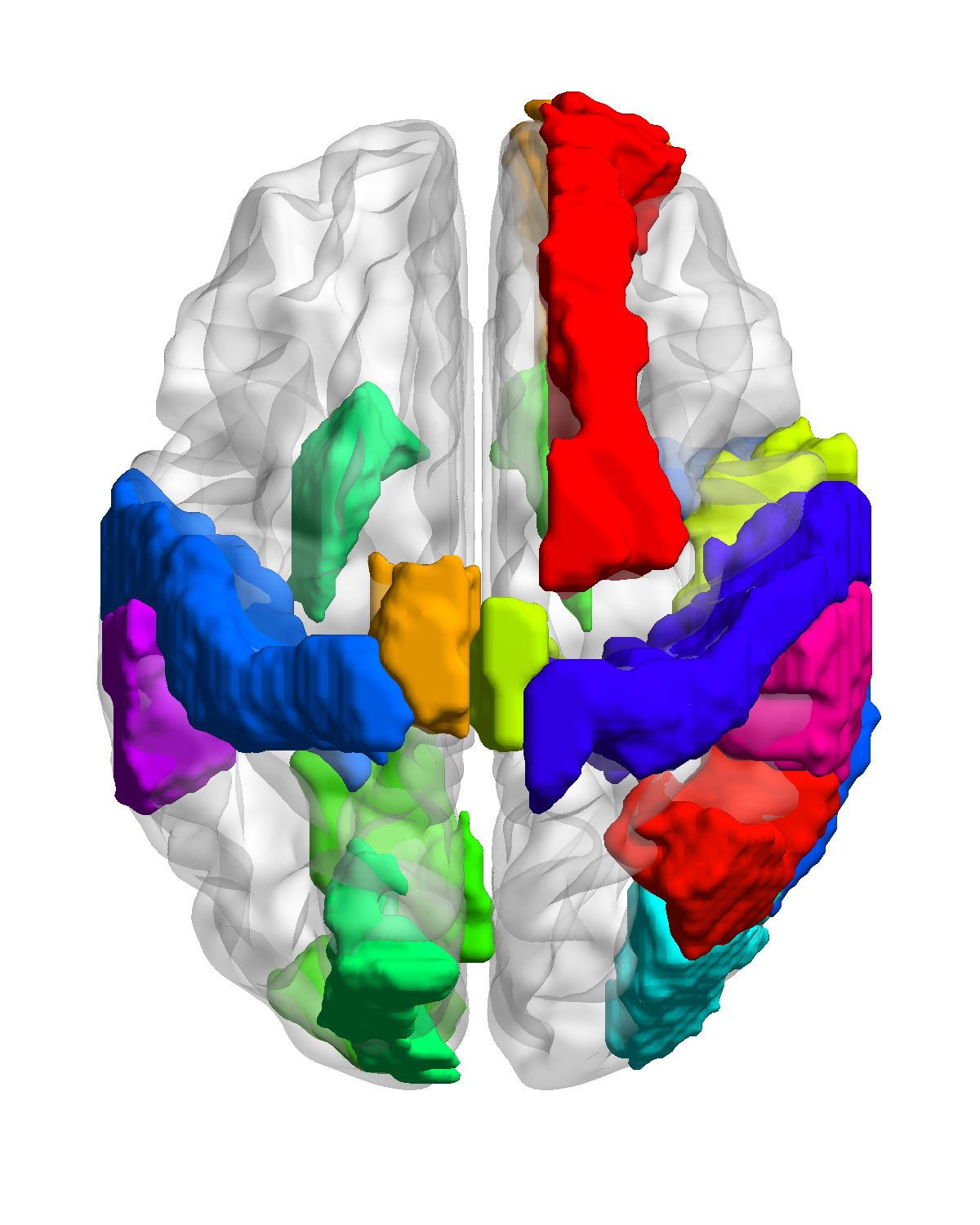}}
	\caption{Visualization of the Alzheimer's disease-related brain regions detected by (a) PANDA, (b) the proposed DGCL and (c) the common regions. Each brain region is represented by a different color, and the same color in the image represents the jointly detected area. From (c), it can be seen that the proposed model has high consistency with the software.}
	\label{fig_11}
\end{figure}

Based on the importance of brain regions, we visualized the brain regions associated with AD generated by PANDA and proposed method, as shown in Fig. \ref{fig_11}. In our four-class classification scenario, 17 ROIs overlapped with the brain regions identified and they show strong consistency in the common regions (TABLE \ref{table7}). These important ROIs include the frontal lobe, the precuneus, the paracentral lobule, the superior frontal gyrus, and the caudate nucleus.

\begin{table*}[!htb]
	\caption{Detection of disease-related brain regions/sub-networks. \label{table7}}
	\centering
	\begin{tabular}{c m{6cm} m{6cm} c}
		\toprule
		\textbf{Comparation}  & \multicolumn{1}{c}{\textbf{Software}}  & \multicolumn{1}{c}{\textbf{Proposed}}  &  \# Common \\
		\midrule
		NC vs. EMCI & 4, 9, 10, 11, 12, 14, 17, 20, 21, 24, 26, 27, 28, 30, 31, 39, 40, 53, 54, 62, 64, 72 & 4, 9, 10, 12, 13, 15, 19, 20, 21, 25, 26, 27, 35, 38, 54, 56, 57, 60, 64, 72, 77, 79 & 10  \\
		NC vs. LMCI & 4, 6, 11, 15, 17, 19, 21, 24, 26, 28, 37, 47, 49, 52, 56, 58, 62, 64, 65, 66, 67, 82 & 6, 17, 18, 19, 20, 21, 24, 28, 45, 47, 49, 51, 52, 56, 57, 58, 59, 63, 64, 65, 66, 67, 81, 82, 89, 90 & 13  \\
		NC vs. AD & 4, 6, 9, 13, 18, 20, 47, 49, 52, 57, 58, 63, 64, 66, 69, 70, 72, 73, 80, 81, 89, 90  & 4, 6, 18, 47, 49, 52, 57, 58, 63, 64, 65, 66, 67, 69, 70, 71, 72, 73, 74, 80, 83, 85, 90 & 17  \\
		HC vs. ASD & 3, 4, 5, 8 & 4, 5, 7, 8 & 3 \\
		\bottomrule
	\end{tabular}
\end{table*}

Most of these regions are consistent with previous research, demonstrating that our model has captured important brain regions associated with AD. According to research in neuroscience and cognitive brain diseases, the widely recognized and highly correlated ROIs in AD/MCI diseases include the hippocampus (HIP. R), inferior frontal gyrus (IFGtriang. L), occipital gyrus (MOG. R), rectus gyrus (REC. L and REC. R), etc \cite{multi, sparse, multiview, mutual}. These ROIs are mainly distributed in the frontal, temporal, and occipital lobes. The frontal lobe is primarily associated with higher-order mental activities, including motor control, language expression, self-awareness, and emotional expression. The temporal lobe is responsible for storing visual and auditory information in memory, and in AD patients, abnormal levels of tau protein have been found in the inferior temporal gyrus \cite{alzheimer2014, abnormal2010}. Overall, the important ROIs and abnormal connections inferred by our model can reflect the major features associated with AD.

\begin{figure}[!h]
	\centering
	\includegraphics[width=3.5in]{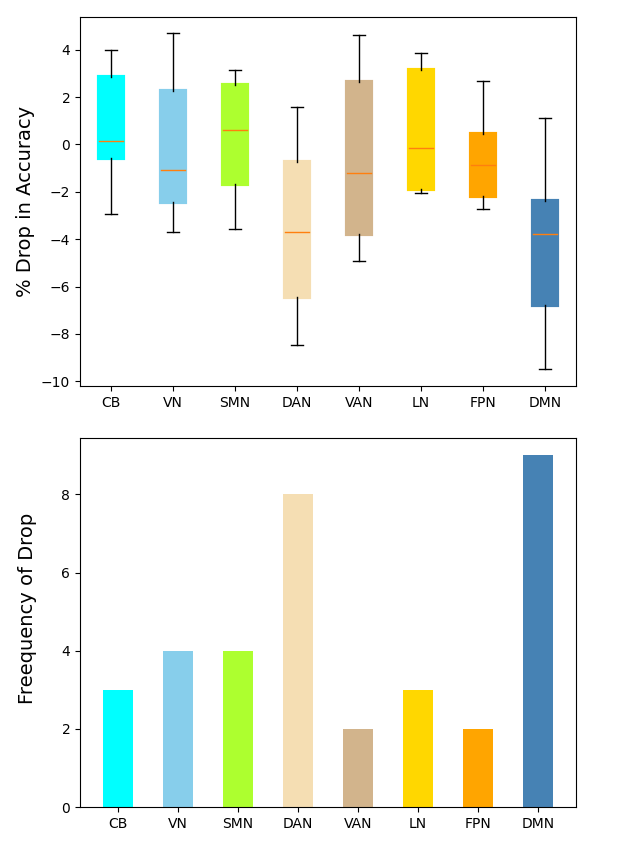}
	\caption{The percentage drop in accuracy (including the median and range) across all sites when each of the sub-networks is occluded in the ablation analysis (top). The frequency of drop in accuracy, i.e. the number of sites where a drop in accuracy is observed, for occlusion of each of the sub-networks in ablation analysis (bottom).}
	\label{fig_12}
\end{figure}

To further validate the importance of brain regions on the autism dataset, we used ablation analysis, specifically, to mask the brain regions of the 8 sub-networks of the imaging data to evaluate the degree of performance degradation. The upper part of Fig. \ref{fig_12} shows the decrease in accuracy when any of the given 8 sub-networks is occluded from the analysis. It can be seen that there are some differences between different brain regions. Therefore, we evaluated the frequency of accuracy decrease, i.e. the number of locations where accuracy decrease is observed in ablation analysis. When each sub network is occluded in the ablation analysis, the percentage of accuracy decrease (median and range) for all stations is also shown in Fig. \ref{fig_12}. In terms of the absolute percentage decrease in frequency and accuracy, both the Dorsal Attention Network (decrease of 8 sites) and the Default Mode Network (decrease of 9 sites) performed outstandingly.

\begin{figure}[!h]
	\centering
	\includegraphics[width=3.4in]{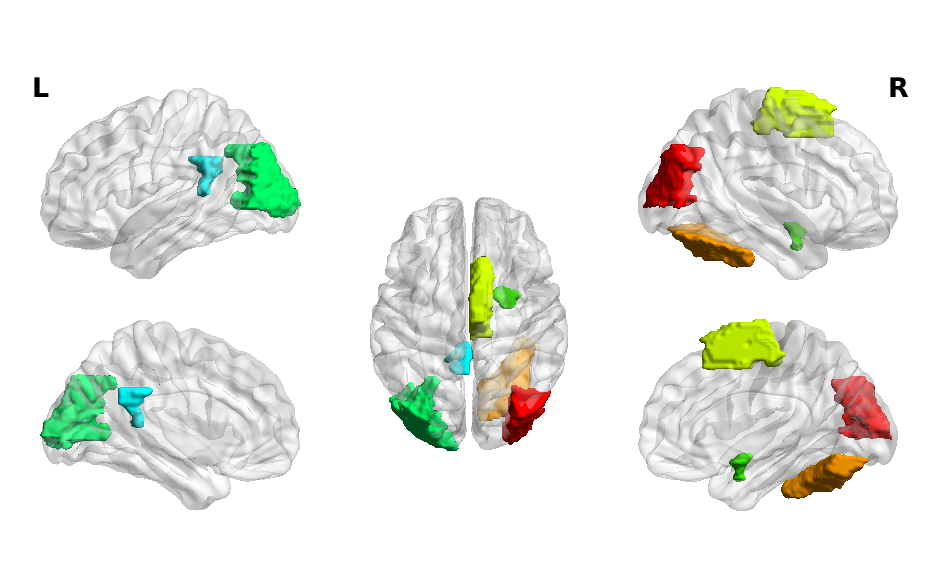}
	\caption{Visualization of the top 5 highest importance of brain region vectors in ASD groups.}
	\label{fig_13}
\end{figure}

% ASD
Distinguish the differences between ASD and HC based on the attention vectors generated by the matrix of the brain network. We study the first row of the learned brain network matrix, which is the correlation coefficient with the corresponding ROI. Use the average cue vector to generate ROI standardized attention scores on correctly classified test data, and we visualized the brain region where TOP5 is located in Fig. \ref{fig_13}. It can be seen that there are significant differences in important brain regions between ASD and HC subjects. In addition, most of the differences in generating vector embeddings occur in DMN and SMN, which is consistent with previous neuroscience literature.

% 5. 结论
\section{Conclusion}
In this paper, we proposed a new brain network reconstruction model based on the diffusion mechanism for aligning brain images and the graph contrastive learning for brain network generation. By using a diffusion-guided BRAM, the model can accurately locate the spatial position of brain regions partitioned by the template and extract important features with spatial information semantics on each brain region. In addition, we use the idea of graph contrastive learning to eliminate individual differences in connectivity unrelated to the disease among populations, and can well retain and strengthen common connections among similar samples, highlighting abnormal brain connections between different categories and improving the efficiency and stability of brain network construction. Although the disease studied in this paper is targeted at AD or ASD, the proposed model can easily be extended to other neurodegenerative diseases. Unlike traditional template-based methods, the proposed model is expected to provide a new deep learning analysis framework for abnormal brain connection detection. Overall, our approach outperforms traditional methods in terms of running efficiency, and accuracy. However, our work has two main limitations: one is that the proposed model still cannot explain the pathogenesis of the detected abnormal brain connections; the other is that the dataset used in the current study is relatively small. In the future, we plan to use other deep learning methods to analyze fMRI data and further study changes in brain function through functional networks. Meanwhile, we may validate the effectiveness of the proposed model on larger medical image datasets such as UK Biobank.

\ifCLASSOPTIONcompsoc
%\else
%  % regular IEEE prefers the singular form
%\fi
%
%This work was supported by the National Key Research and Development Program of China under Grant 2023YFC2506902, National Natural Science Foundations of China under Grant 62172403, U22A2024, and 62271328.

\end{document}